# Energy stability of supercontinuum via femtosecond filamentation in sapphire

CHEN JIUCHENG[1], XIAO HENGYUAN[1], ZHANG TIANLIANG[1], DING SIQIN[1], AND HUA JIANFEI[1],*

[1]*Department of Engineering Physics, Tsinghua University, Beijing 100084, China*
*jfhua@tsinghua.edu.cn*

**Abstract:** The energy stability of supercontinuum (SC) significantly impacts its applications. To achieve the most stable SC, we systematically investigated how input pulse energy, numerical aperture (NA), and crystal thickness affect the energy stability of SC generated by femtosecond filamentation in sapphire. Our findings reveal that the SC energy does not always increase monotonically with input energy for different NA and thicknesses. This phenomenon occurs because, when the input pulse energy just exceeds the filamentation threshold, the pulse splitting structure and spectrum are still rapidly evolving. To generate a more stable SC, the numerical aperture and crystal thickness must be carefully coordinated to prevent this rapid evolution from occurring within the crystal.

## 1. Introduction

Supercontinuum (SC) generation via femtosecond filamentation in bulk materials has become an efficient method for producing broadband coherent light sources, with widespread applications in ultrafast spectroscopy and nonlinear optics [1]. Notable implementations include its use as a probe source in transient absorption spectroscopy [2] and multidimensional optical spectroscopy [3]，as well as a seed for optical parametric amplifiers (OPA) [4-6] and optical parametric chirped pulse amplification (OPCPA) systems [7-9].

Filamentation is a complex nonlinear process involving self-focusing, optical Kerr effect, ionization, plasma defocusing and material dispersion [1]. As a laser pulse propagates within a nonlinear crystal, self-focusing initially induces spatiotemporal compression, leading to a progressively increase in optical density. This intensification continues until the catastrophic collapse at the nonlinear focus is arrested by group velocity dispersion (GVD) and plasma generated by ionization [10]. In the regime of normal GVD, typical pulse splitting occurs at the nonlinear focus [11,12], accompanied by explosive broadening of the spectrum. The leading sub-pulse contains the red-shifted SC components, while the trailing sub-pulse contains the blue-shifted components [13].

Despite the complexity of filamentation, its experimental implementation is straightforward, requiring only the focusing of a μJ-level femtosecond laser pulse into a bulk condensed medium [14]. For a given combination of laser source and crystal material, filamentation is predominantly governed by three key parameters: the energy of the driving laser pulse, the numerical aperture (NA), which determines the focal spot size, and the crystal thickness. The energy of the driving laser pulse must be carefully adjusted to maintain a single filamentation state，as multiple filamentations can cause spectral modulation due to interference between spectra produced by a series of pulse splittings [15]. Variations in NA has a significant impact on the infrared extent of SC, while the maximum blue-shifted spectrum is almost unaffected [16]. Additionally, a high NA increases the risk of crystal damage [17]. In the case of picosecond filamentation, it has been experimentally found that a stable SC can be produced via exceptionally long YAG crystal thickness and loose focusing conditions [9,18].

To optimize SC energy stability, we conducted a comprehensive study of the effects of driving pulse energy, focal spot size and crystal thickness on SC energy stability through both simulations and experiments using sapphire crystals. Our study found that, in the single filamentation regime, SC energy does not increase monotonically with input pulse energy but instead initially decreases before subsequently increasing. Numerical simulations reveal that the decrease in SC energy occurs when the input pulse energy is just above the filamentation threshold, during which the temporal and spectral shapes of the pulse are still rapidly evolving. Once the pulse splitting stabilizes into an asymmetrical state, the SC energy increases monotonically. Furthermore, as the input pulse energy increases, the spatial distance between the first and secondary pulse splittings shortens, ultimately leading to the disappearance of the SC energy plateau.

## 2. Numerical Simulations

The filamentation of femtosecond pulse in sapphire crystal is simulated by solving the nonlinear Schrödinger equation [19] in cylindrical geometry

$$\frac{\partial \tilde{A}}{\partial \zeta} = \frac{i}{2k_0}\Delta_\perp \tilde{A} + \frac{ik_0''}{2}\Omega^2 \tilde{A} + \frac{i}{2n_0}\frac{\omega_0}{c}\frac{\tilde{P}}{\varepsilon_0}, \tag{1}$$

where $\tilde{A}$ is the complex envelope of the laser pulse in the spectral domain, $k_0$ is the wave number, $k_0''$ is the group velocity dispersion at the central wavelength, and $\Omega$ equals to $\omega - \omega_0$. The nonlinear polarization $P$ includes optical Kerr effect and ionization [20,21]

$$P = \frac{n_2 n_0^2 \varepsilon_0}{\mu_0 c}|A|^2 A + \frac{i\varepsilon_0 c n_0}{\omega}E_g \frac{W}{I}\left(1-\frac{\rho}{\rho_{nt}}\right)A + \frac{i\varepsilon_0 c n(\omega)}{\omega}\sigma(1+i\omega\tau_{col})\rho A, \tag{2}$$

where $n_2$ is the nonlinear refractive index, $E_g$ is the bandgap of the crystal, $\rho$ is the density of free electron density, $\rho_{nt}$ is density of neutral atoms, $\sigma$ is cross-section for inverse Bremsstrahlung, and $\tau_{col}$ is the effective collision time. The ionization rate $W$ is calculated from Keldysh's theory [22], with the reduced electron-hole mass set to 0.3 $m_e$. The free electron density is obtained by solving the rate equation

$$\frac{\partial \rho}{\partial t} = W(|A|)\left(1-\frac{\rho}{\rho_{nt}}\right) + \frac{\sigma(\omega_0)}{E_g}\rho I\left(1-\frac{\rho}{\rho_{nt}}\right) - \frac{\rho}{\tau_{rec}}, \tag{3}$$

where $\tau_{rec}$ is the electron recombination time. All parameters for sapphire crystals are taken from Ref.[16].

Fig. 1(a) and (b) shows the evolution of the on-axis temporal envelope and angularly averaged spectrum of a 34 fs pulse with an input focal spot size of 70 μm (FWHM) in a 7 mm sapphire crystal. The input pulse energy is set to 800 nJ, which is above the filamentation threshold but insufficient for the pulse to refocus inside the crystal. The pulse splitting is accompanied by instantaneous spectral broadening, and the temporal profiles of the two sub-pulses exhibit asymmetry during their propagation along the crystal. Fig. 1(d) shows the output SC spectrum, with the 500-600 nm band selected as the region of interest. This band represents a typical flat plateau on the blue-shifted side of SC generated by 800 nm laser pulses from a Ti:sapphire laser [14], and the blue-shifted side of a SC is often used to seed OPCPA systems [7]. Fig. 1(c) shows the SC (500-600 nm) energy, beam diameter, and peak intensity as a function of crystal length. The SC energy increases only during periods of high laser intensity. We performed a series of simulations with varying input focal spot sizes and energies at a fixed 7 mm crystal length. All simulations were conducted under single filamentation conditions. Fig. 2 shows the SC energy as a function of input pulse energy. A plateau in the SC energy curves is always present, regardless of the input spot size. However, the plateau is not always monotonic for smaller input spot sizes. Specifically, for a 20 μm focal spot size, the SC energy decreases initially before increasing. An almost flat plateau can be achieved with an optimal input spot size. For larger focal spot sizes (e.g., 80 μm), more input energy is required to reach the plateau, and the plateau becomes shorter.

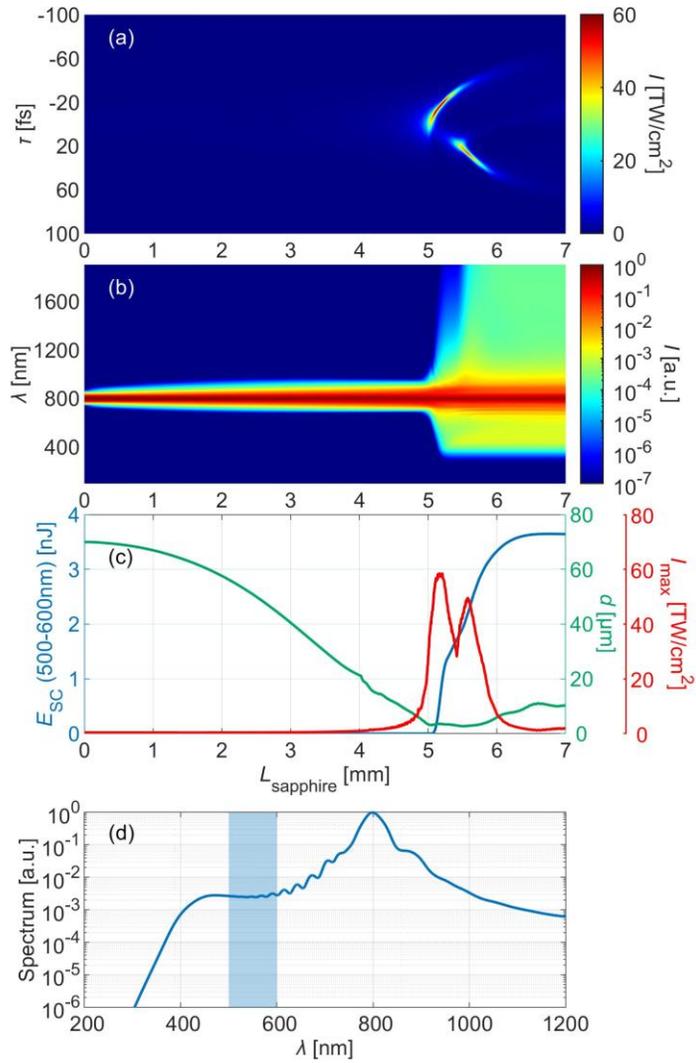

Fig. 1. Simulation results of the propagation of a 34 fs, 70 μm, 800 nJ pulse in a 7 mm sapphire crystal: (a) on-axis temporal envelope evolution, (b) angularly averaged spectrum, (c) SC energy (500-600nm), beam diameter (FWHM), and maximum intensity as a function of crystal length, (d) the SC spectrum at 7 mm length, with the shaded area showing the 500-600nm band of interest.

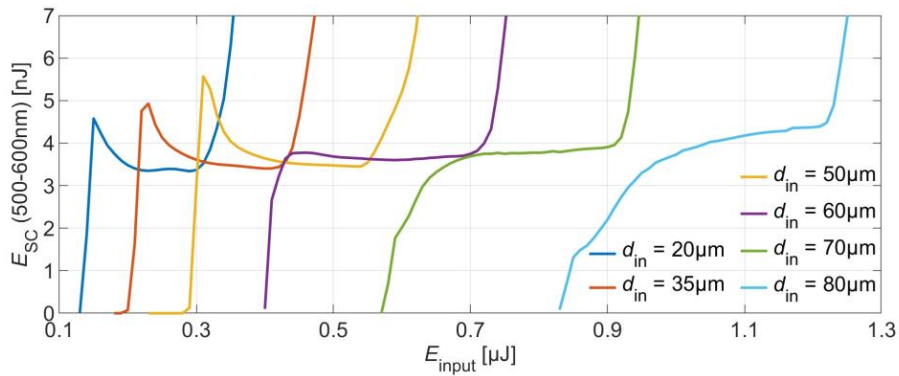

Fig. 2. The SC energy (500-600nm) versus input pulse energy with different input focal sizes (FWHM).

## 3. Experimental Results

To validate the simulation results, we performed experimental studies using a Ti:sapphire regenerative amplifier operating at a repetition rate of 1 kHz and a pulse duration of 34 fs. The amplifier delivered 3mJ of energy, which was attenuated to less than 100 µJ using beam splitters and adjusted with a λ/2 wave plate and polarizing beam splitter. The pulse was focused onto the front face of the sapphire crystal using an $f_1$ lens with a focal length of 200 mm, and the generated SC was collected using an $f_2$ lens with a focal length of 50 mm. A long pass filter (FELH0500, Thorlabs) and a short pass filter (FESH0600, Thorlabs) were used to select the 500-600 nm band of interest. The focal spot size was adjusted using an iris placed in front of the $f_1$ lens. The numerical aperture and focal spot size were calculated using the formula:

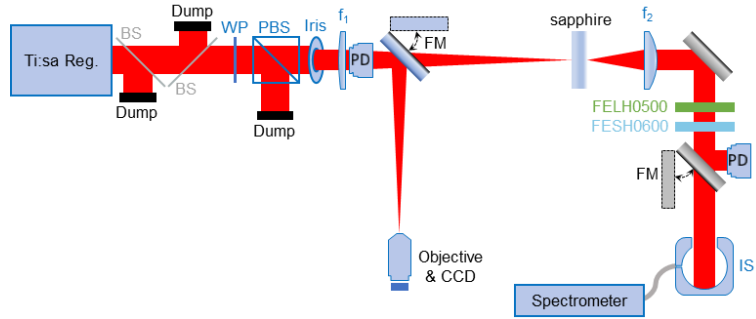

Fig. 3. Experiment set-up. BS: beamsplitter, WP: λ/2 wave plate, PBS: polarizing beamsplitter, PD: photodiode power meter, FM: flip mirror, IS: integrating sphere.

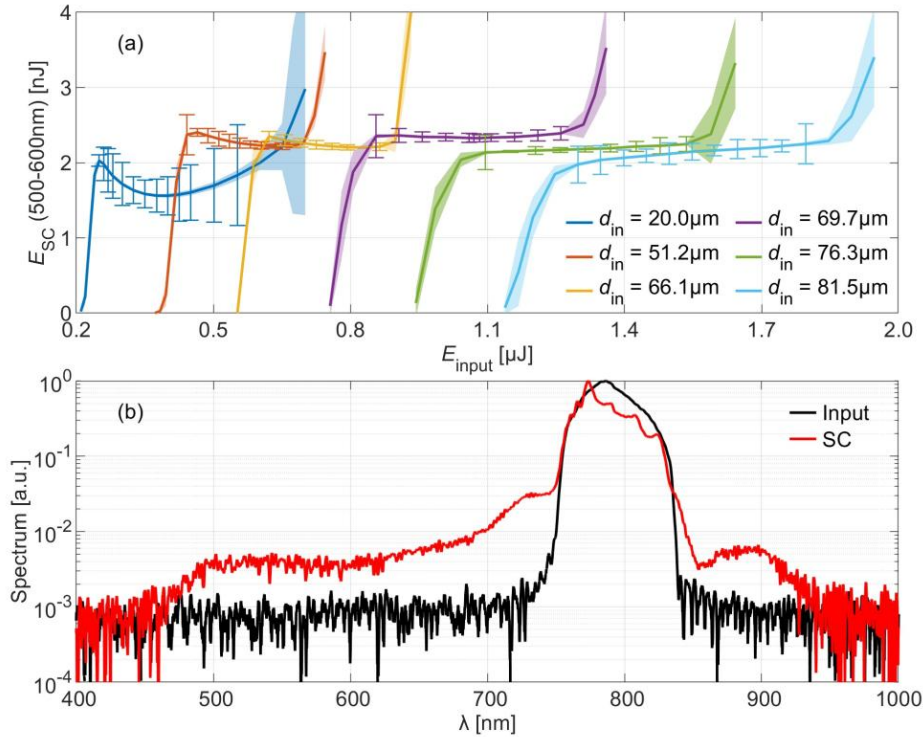

Fig. 4. Experimental results: (a) SC energy (500-600nm) versus input pulse energy with different input focal sizes (FWHM); (b) the spectra of input pulse and SC, measured at 69.7 µm focal size and 1.07 µJ input pulse energy.

$d_{in} = \sqrt{2\ln 2} M^2 \lambda / \pi \text{NA}$. A photodiode power meter (PD300-UV, Ophir) was placed either right after $f_1$ to measure the drive pulse energy, or after the two filters to measure the SC energy. The focal spot size was measured using an objective and CCD, and the SC spectrum was collected using an integrating sphere (2P3/M, Thorlabs) and measured with a spectrometer (USB4000, Ocean Optics). The two filters were also removed when measuring the SC spectrum.

Fig. 4(a) shows the experimentally measured SC energy as a function of input pulse energy. The shaded area represents 10 times the standard deviation (STD) values, while the error bars represent 100 times the STD values of the plateau sections. All the data were measured for one minute before being recorded. The experimental results confirm that the numerical aperture has a significant effect on the SC energy plateau. By choosing an appropriate focal spot size, a nearly flat plateau can be achieved, resulting in the most stable SC. The red curve in Fig. 4(b) represents measured typical spectrum of SC, which was generated with a 69.7 μm focal spot size and 1.07 μJ input pulse energy. A plateau is observed in the 500-600 nm band.

## 4. Discussion

The decrease in SC energy is also indicated the spectra. Fig. 5(a) shows the simulated SC spectra for a 20 μm focal spot size with different input pulse energies, corresponding to the descending section of the SC energy curve. As the SC spectrum expands from 700 nm to shorter wavelengths, its energy gradually decreases. For a 70 μm focal spot size, the energy in the 400-700 nm band increases to a certain value and then stabilizes, as shown in Fig. 5(b). This evolution of SC spectra with input pulse energies at different focal spot sizes was also observed experimentally, as shown in Fig. 5(c) and (d), corresponding to focal spot sizes of 20.0 μm and 76.3 μm, respectively. The experimental spectra presented in (c) and (d) have been noise-filtered.

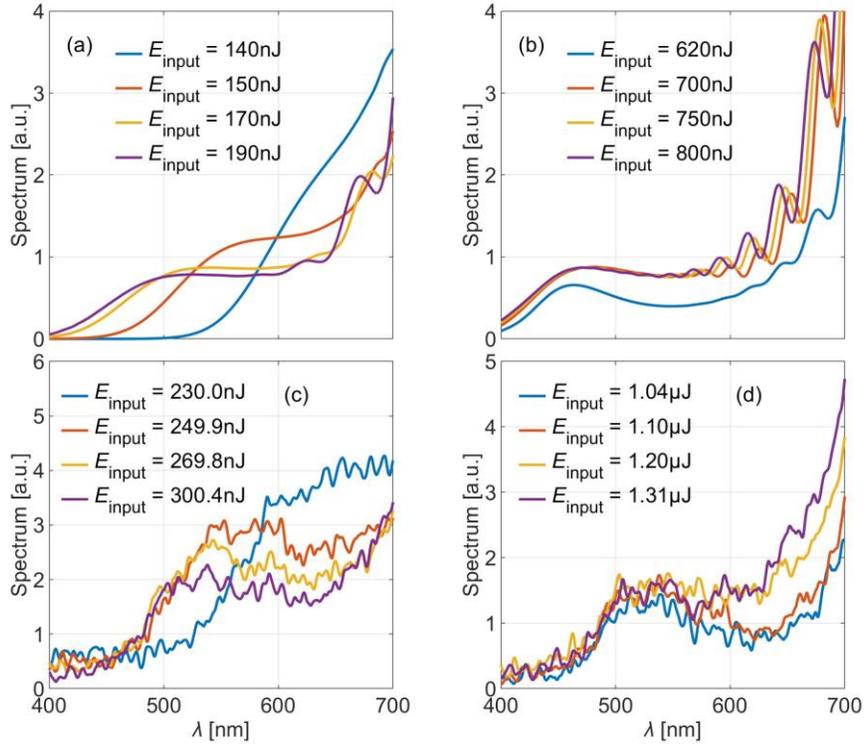

Fig. 5. SC spectra with different input pulse energies. (a) and (b) correspond to simulation results for focal spot sizes of 20 μm and 70 μm. (c) and (d) correspond to experimental results for focal spot sizes of 20.0 μm and 76.3 μm.

We attribute the decrease in SC energy to the input pulse energy being slightly above the filamentation threshold. For example, while Fig. 2 shows that SC energy increases monotonically for a 70 μm focal spot size, it can decrease with input pulse energy if a longer sapphire crystal is used, as shown in Fig. 6(j). The curves in Fig. 6(j) correspond to different crystal thicknesses. Fig. 6(a)-(i) show the evolution of on-axis temporal envelope (subscript 1) and the angularly averaged spectrum (subscript 2) for various input pulse energies. The corresponding energies are marked with black dashed lines in Fig. 6(j) in order. In the decreasing portion (0.45-0.65 μJ) of the SC energy curve, the pulse splitting evolves from a symmetrical shape into an asymmetrical one. Fig. 6(k) shows that the maximum peak intensity increases rapidly during this portion, while the minimum pulse diameter decreases. Once the typical asymmetrical pulse splitting is formed, the maximum peak intensity increases slowly with input pulse energy, and the minimum pulse diameter remains nearly constant. After the SC energy reaches its local minimum, further increases in input pulse energy do not significantly change the pulse splitting shape but advance its position within the crystal. Fig. 6(f)-(i) show that the position of the first pulse splitting changes less as the input energy increases, as the nonlinear focus is approximately inversely proportional to the input pulse energy [23]. However, increasing the input energy rapidly advances the position of the secondary pulse splitting until it catches up with the first pulse splitting, causing the SC energy plateau to disappear, as shown in Fig. 6(j) for a 5 mm crystal thickness. Fig. 6(l) shows the SC energy in the 500-600nm band as a function of crystal thickness and input pulse energy. The colormap is adjusted for a better indication of the single filamentation regime. It is clear that the thickness range for single filamentation shrinks as the input pulse energy increases.

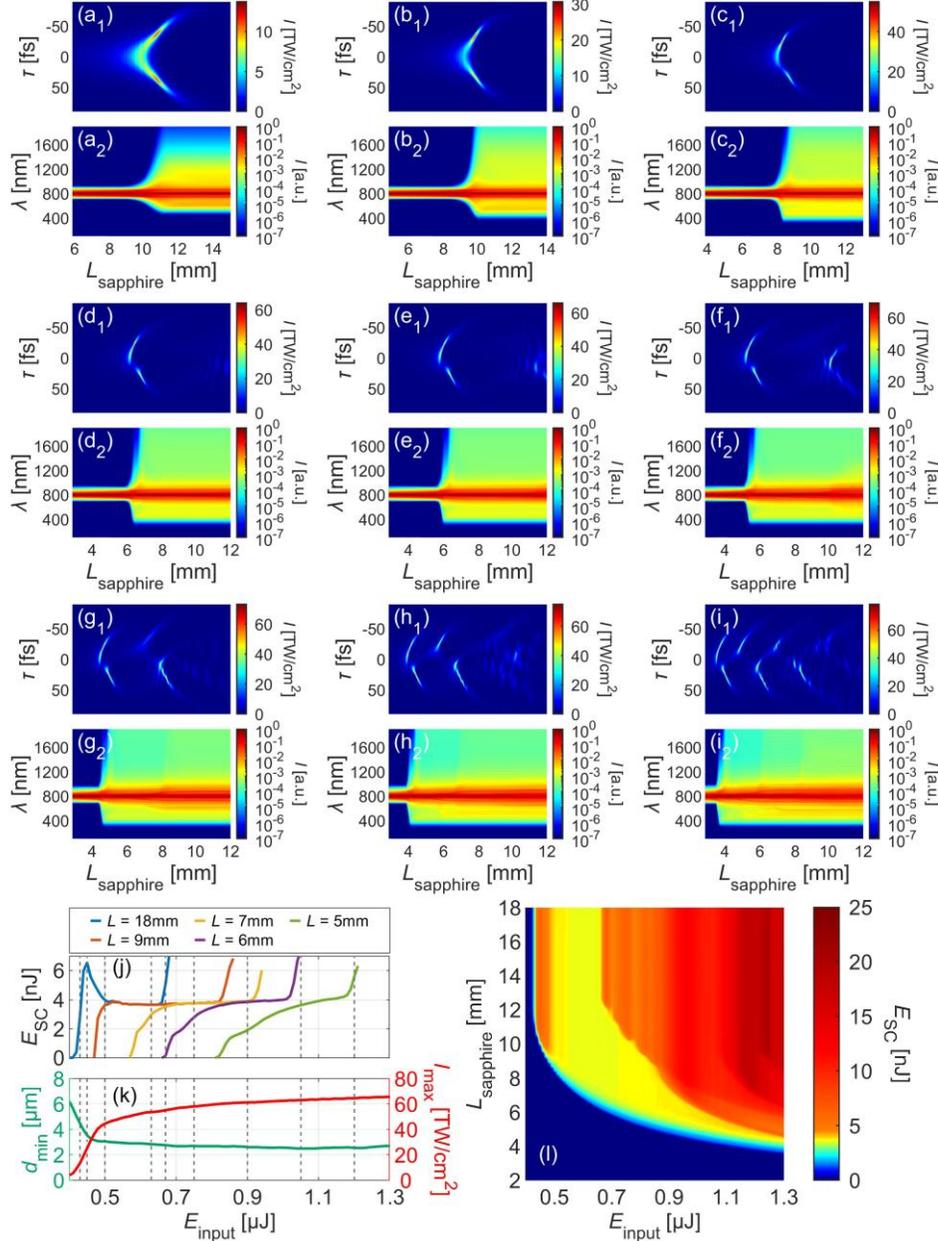

Fig. 6. Simulation results of propagation of a 34 fs, 70 μm pulse in a 18 mm sapphire crystal: (a)-(i), evolution of on-axis temporal envelope and angularly averaged spectrum with various input energy, (j) SC energy (500-600nm) and (k) minimum beam diameter (FWHM) and maximum intensity as a function of crystal length, (l) SC energy (500-600nm) as a function of crystal length and input pulse energy.

These conclusions are validated by experimental results obtained with different crystal thicknesses. Fig. 7(a) shows the SC energy as a function of input energy for a 21.5 μm focal spot size. The curves for 7 mm, 6 mm, 5 mm, and 4 mm crystal thicknesses overlap almost completely, indicating that filamentation takes place at a crystal length of less than 4 mm, thus the decrease in SC energy can be observed. Fig. 7(b) shows the experimental result for a 76.3 μm focal spot size. In this case, using a 6 mm crystal yield an SC energy curve that extends the plateau, but further reducing the crystal thickness to 5 mm results in a curve which does not overlap with the 6 mm curve at all, and does not show s distinct plateau.

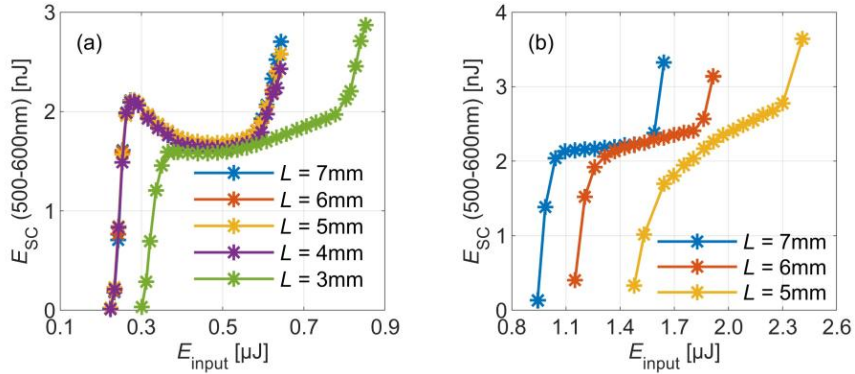

Fig. 7. Experimental results of the SC energy (500-600nm) as a function of input pulse energy with different crystal length measured in two different cases of focal spot sizes: (a) 21.5 μm, (b) 76.3 μm.

Therefore, the focal spot size and crystal thickness must be carefully selected to ensure that symmetrical pulse splitting due to input pulse energy just above the filamentation threshold does not occur within the crystal. Additionally, the secondary pulse splitting should not be too close to the first pulse splitting, as this would cause the SC energy plateau to disappear. Furthermore, since both self-focusing and pulse splitting require a certain distance (1-2 mm) within the crystal, the crystal thickness should not be less than this.

Finally, the focal spot size should not be too small, as this would result in a very shorter SC energy plateau. The connected plateaus in the Fig. 6(j) from 18 mm to 6 mm correspond to the complete SC energy curve in the single filamentation regime for a 70 μm focal spot size. Increasing the thickness to 19 mm results in an SC energy curve that coincides with the 18 mm curve, while reducing the thickness to 5 mm causes the plateau to disappear. Fig. 8 shows the complete SC energy curves in the single filamentation regime as a function of input energy for different focal spot size. The figure shows that the SC energy plateau becomes shorter as the input focal size decreases, which is why the energy STD remains large even when there is a local minimum in the SC energy plateau for a 20.0 μm focal spot size, as shown in Fig. 4(b).

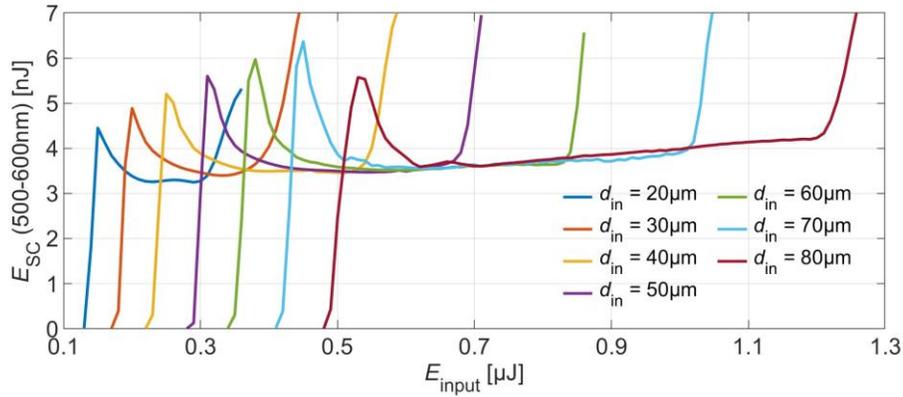

Fig. 8 The energy of SC through one single filamentation as a function of input energy in the case of different focal spot sizes.

Therefore, the focal spot size should not be too small to ensure that a proper plateau can be intercepted from the complete SC energy curve by choosing an appropriate crystal thickness.

## 5. Conclusions

In summary, we conducted a systematic study of the effects of focal spot size, which is determined by NA, crystal thickness, and input pulse energy on the energy stability of SC generated via filamentation induced by femtosecond laser pulse. Simulations revealed that pulse splitting exhibits a symmetric shape when the laser pulse is still rapidly evolving due to input energy just above the filamentation threshold. Increasing the input energy further allows the formation of a stable, asymmetrical pulse splitting. However, the position of the secondary pulse splitting advances much faster than the first pulse splitting, causing the SC energy plateau to disappear if the two pulse splittings become too close. One can generate a SC that is as stable as possible by carefully selecting these three parameters. These findings are very useful from a practical point of view.

**References**

[1] Dubietis, A., Tamošauskas, G., Šuminas, R., Jukna, V., & Couairon, A. Ultrafast supercontinuum generation in bulk condensed media[J]. Lithuanian Journal of Physics, 2017, 57(3), 113-157.
[2] Wang, J., Zhang, Y., Shen, H., Jiang, Y., & Wang, Z. Spectral stability of supercontinuum generation in condensed mediums[J]. Optical Engineering, 2017, 56(7).
[3] Song, Y., Konar, A., Sechrist, R., Roy, V. P., Duan, R., Dziurgot, J., . . . Ogilvie, J. P. Multispectral multidimensional spectrometer spanning the ultraviolet to the mid-infrared[J]. Review of Scientific Instruments, 2019, 90(1).
[4] Wilson, K. R., & Yakovlev, V. V. Ultrafast rainbow: Tunable ultrashort pulses from a solid-state kilohertz system[J]. Journal of the Optical Society of America B-Optical Physics, 1997, 14(2), 444-448.
[5] Baltuska, A., Fuji, T., & Kobayashi, T. Visible pulse compression to 4 fs by optical parametric amplification and programmable dispersion control[J]. Optics Letters, 2002, 27(5), 306-308.
[6] Cerullo, G., & De Silvestri, S. Ultrafast optical parametric amplifiers[J]. Review of Scientific Instruments, 2003, 74(1), 1-18.
[7] Budriūnas, R., Stanislauskas, T., & Varanavičius, A. Passively CEP-stabilized frontend for few cycle terawatt OPCPA system[J]. Journal of Optics, 2015, 17(9).

[8] Shao, B., Li, Y., Peng, Y., Wang, P., Qian, J., Leng, Y., & Li, R. Broad-bandwidth high-temporal-contrast carrier-envelope-phase-stabilized laser seed for 100 PW lasers[J]. Optics Letters, 2020, 45(8), 2215-2218.

[9] Indra, L., Batysta, F., Hříbek, P., Novák, J., Hubka, Z., Green, J. T., . . . Rus, B. Picosecond pulse generated supercontinuum as a stable seed for OPCPA[J]. Optics Letters, 2017, 42(4), 843-846.

[10] Faccio, D., Clerici, M., Averchi, A., Lotti, A., Jedrkiewicz, O., Dubietis, A., . . . Di Trapani, P. Few-cycle laser-pulse collapse in Kerr media: The role of group-velocity dispersion and X-wave formation[J]. Physical Review A, 2008, 78(3).

[11] Rothenberg, J. E. Pulse splitting during self-focusing in normally dispersive media[J]. Optics Letters, 1992, 17(8), 583-585.

[12] Ranka, J. K., Schirmer, R. W., & Gaeta, A. L. Observation of pulse splitting in nonlinear dispersive media[J]. Physical Review Letters, 1996, 77(18), 3783-3786.

[13] Gaeta, A. L. (2009). Spatial and Temporal Dynamics of Collapsing Ultrashort Laser Pulses. In R. W. Boyd, S. G. Lukishova, & Y. R. Shen (Eds.), *Self-Focusing: Past and Present: Fundamentals and Prospects* (Vol. 114, pp. 399-411).

[14] Bradler, M., Baum, P., & Riedle, E. Femtosecond continuum generation in bulk laser host materials with sub-µJ pump pulses[J]. Applied Physics B, 2009, 97(3), 561-574.

[15] Jarnac, A., Tamosauskas, G., Majus, D., Houard, A., Mysyrowicz, A., Couairon, A., & Dubietis, A. Whole life cycle of femtosecond ultraviolet filaments in water[J]. Physical Review A, 2014, 89(3).

[16] Jukna, V., Galinis, J., Tamosauskas, G., Majus, D., & Dubietis, A. Infrared extension of femtosecond supercontinuum generated by filamentation in solid-state media[J]. Applied Physics B-Lasers and Optics, 2014, 116(2), 477-483.

[17] Ashcom, J. B., Gattass, R. R., Schaffer, C. B., & Mazur, E. Numerical aperture dependence of damage and supercontinuum generation from femtosecond laser pulses in bulk fused silica[J]. Journal of the Optical Society of America B-Optical Physics, 2006, 23(11), 2317-2322.

[18] Roškot, L., Novák, O., Csanaková, B., Mužík, J., Smrž, M., Jelínek, M., & Mocek, T. Importance of crystal length on the stability of a picosecond supercontinuum generated in undoped YAG[J]. Journal of the Optical Society of America B, 2023, 40(6).

[19] Couairon, A., Brambilla, E., Corti, T., Majus, D., de J. Ramírez-Góngora, O., & Kolesik, M. Practitioner's guide to laser pulse propagation models and simulation[J]. The European Physical Journal Special Topics, 2011, 199(1), 5-76.

[20] Galinis, J., Tamošauskas, G., Gražulevičiūtė, I., Keblytė, E., Jukna, V., & Dubietis, A. Filamentation and supercontinuum generation in solid-state dielectric media with picosecond laser pulses[J]. Physical Review A, 2015, 92(3).

[21] Spacek, A., Indra, L., Batysta, F., Hribek, P., Green, J. T., Novak, J., . . . Rus, B. Stability mechanism of picosecond supercontinuum in YAG[J]. Opt Express, 2020, 28(14), 20205-20214.

[22] Keldysh, L. V. Ionization in the field of a strong electromagnetic wave[J]. Soviet Physics JETP, 1965, 20(5), 1307-1314.

[23] Marburger, J. H. Self-focusing: theory[J]. Progress in Quantum Electronics, 1975, 4(1), 35-110.